# Expanding the Scope of Computational Thinking in Artificial Intelligence for K-12 Education

Yasmin B. Kafai, University of Pennsylvania, kafai@upenn.edu
Shuchi Grover, Looking Glass Ventures, shuchi.stanford@gmail.com

ABSTRACT
The introduction of generative artificial intelligence applications to the public has led to heated discussions about its potential impacts and risks for K-12 education. One particular challenge has been to decide what students should learn about AI, and how this relates to computational thinking, which has served as an umbrella for promoting and introducing computing education in schools. In this paper, we situate in which ways we should expand computational thinking to include artificial intelligence and machine learning technologies. Furthermore, we discuss how these efforts can be informed by lessons learned from the last decade in designing instructional programs, integrating computing with other subjects, and addressing issues of algorithmic bias and justice in teaching computing in schools.

INTRODUCTION

The public release of OpenAI's ChatGPT in November 2022 launched an unprecedented frenzy of discussions about the impact of artificial intelligence and machine learning (AI/ML)[1] on K-12 education. Conversations have ranged from highlighting the opportunities for personalized learning, automatic feedback, and grading to raising issues about setbacks in learning and cheating in homework and exams, as well as debates and commentaries on teaching and learning about AI/ML (US Department of Education Report, 2023). In particular, evidence of algorithmic bias and harm have been noted as critical issues in educational applications (e.g., Broussard, 2018). While applications such as recommender systems or chatbots found in search engines and social media have been in use for more than decade, the accelerated pace with which new AI applications are being released has made it difficult for schools (and everyone else for that matter) to keep up with developments and provide recommendations about appropriate and relevant uses of AI/ML for student learning and teaching.

With the pace at which AI/ML is advancing and being used, K-12 education in general, but CS education in particular, now finds itself at a crossroads (Grover, 2024). Even amid continuing debate about appropriate uses of more "traditional" AI by teachers and students, the conversation has now rapidly shifted to what teachers and students should know about AI/ML. Current national frameworks and standards in K-12 education—and even K-12 CS education—do not explicitly address AI/ML concepts and practices such as neural nets and algorithmic bias (e.g., Tedre et al., 2021). Different proposals of what exactly every student would need to know about AI have been developed (e.g., Long & Magerko, 2020; Touretzky et al., 2019), and new definitions of computational thinking (CT) for designing AI models are emerging (Tedre et al., 2021). Likewise, teachers across all subjects and grades are equally underprepared to integrate AI/ML applications into their classroom practice (Olari & Romeike, 2021). Finally, numerous classroom materials have been developed, ranging from short Hour of Code introductions to week-long curricula (e.g., see Druga et al., 2022), but there is no consensus of where the learning about AI/ML should be situated within an already crowded school curriculum. While some see AI/ML as part of computing education, which is slowly making its way into school classrooms (DeLyser et al., 2018), others would like to see it as part of the general curriculum. In short, barring a few efforts to lend coherence to the discourse (e.g., Grover 2024), current conversations about AI/ML are currently a chaotic cacophony with little explicit framing for making sense of how to integrate AI into CS and non-CS settings.

---

[1] Although AI is a broad field that began in the 1950s, it is only machine learning—the most recent and popular application of AI—that is the current focus of AI initiatives in education, hence our use of AI/ML.

With computing as a field constantly evolving, what it means to be computationally literate is a moving target, putting educational researchers and practitioners alike in an unusual position. In this paper, we address the need for guiding frameworks that accommodate AI and offer our take on how the education community, in particular K-12 CS education, can move forward. We propose to lend coherence to the discourse by situating the conversation about AI/ML in K-12 education in the context of computational thinking (Wing, 2006), a construct that has provided instrumental leverage to K-12 CS education over the last decade. Computational thinking (hereafter: CT) has been broadly defined as "involving solving problems, designing systems, and understanding human behavior that draws on concepts fundamental to computing" (Wing, 2006, p. 33). In practice, CT is the umbrella goal under which CS education has been introduced in schools and in teacher preparation around the globe (Hsu, Irie, & Ching, 2019). Using it to frame the conversation thus has practical value. Additionally, evolving conceptualizations of CT, and in particular the most recent framing by Kafai and Proctor (2022) can readily be expanded to include skills and competencies for computational problem-solving with AI/ML. A second reason for situating AI/ML education within CT is that it can greatly benefit from the lessons learned in developing over decades a broad knowledge base in computing education. In the following sections, we offer an expanded definition of CT for AI/ML and outline areas of focus for K-12 research and teaching.

EXPANDING THE SCOPE OF COMPUTATIONAL THINKING

Since CT's arrival in K-12 Education (Grover & Pea, 2013), it has become the foundational concept for promoting computer science education in schools. Over the last decade, it has guided efforts to develop a national framework (K12 Framework, 2016), establish teacher certification standards and teacher education programs, set up CS courses in schools, and count them towards graduation requirements (e.g., deLyser, Goode, Guzdial, Kafai and Yadav, 2018). The need to include these socio-cultural and socio-political dimensions in addition to technical concepts and practices of CT led Kafai and Proctor (2022) to call for the development of computational literacies as part of CT learning in K-12 education with the overall goal has been to help students become computationally literate, moving from a narrow technical understanding to include also "values, biases, and histories embedded in computational technologies and cultures which run on computers" (p. 147). Computational literacies encompass phenomena at scales from the individual to the societal, as well as connections between these phenomena and the media that supports and shapes them (see Figure 1). Computational literacies then cast a wider net and can focus not just on technical components such as programming, data, algorithms, and information but also attend to the social, cultural, and critical issues prevalent in computing (Tedre, Simon, & Malmi, 2018) that can leverage insights we have gained from decades of efforts getting computing into school.

In the following sections, we outline how an expanded scope of CT can address AI/ML concepts and practices while also directing our attention to the growing presence of CT across the school curriculum and attending to the equally important critical, ethical, and justice-centered issues in CS as well as non-CS classrooms.

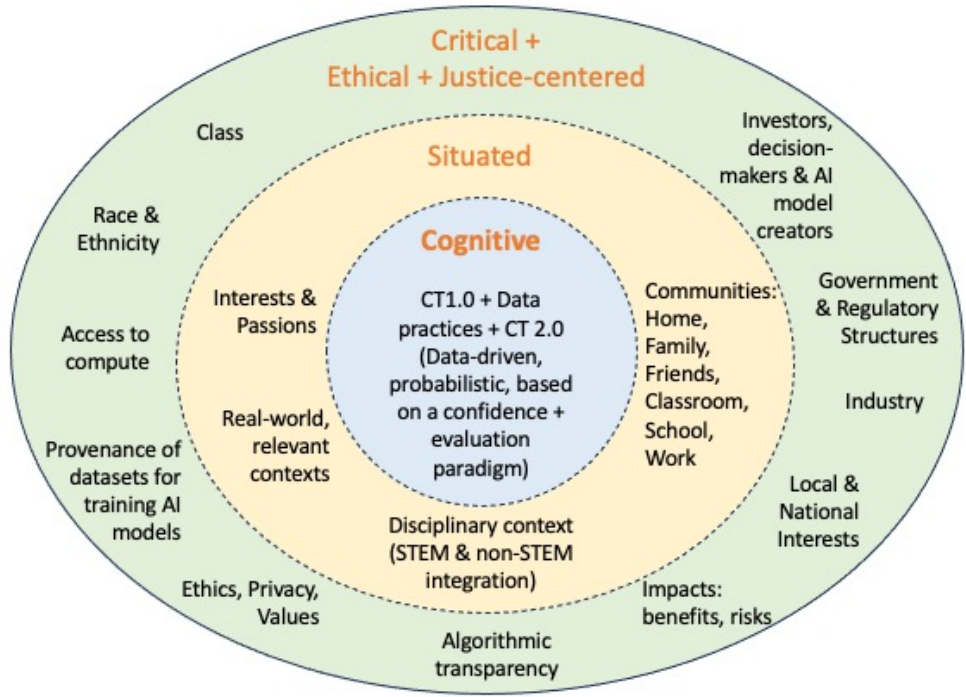

**Figure 1.** Expanded View of Computational Thinking: CT+ based on Kafai & Proctor (2022)

Cognitive

The cognitive aspect of CT is in the vein of what was first proposed in Wing's (2006) timely and influential paper. This earliest version of CT in K-12, or "CT 1.0" (Tedre et al., 2021), has become widely recognized and encompasses conceptual and problem-solving strategies which include components such as algorithmic and logical thinking, representation and abstraction, generalization and pattern recognition, problem decomposition, debugging, systematic error correction, and evaluation. Subsequently expounded on the K-12 education community by definitions that made CT education more tractable and actionable for curriculum designers, researchers, and teachers (e.g., Barr & Stephenson, 2011; Grover & Pea, 2013). These articles collectively emphasized CT 1.0's components as important in K-12 CS education as a composite set of problem-solving approaches that transcend programming environments and that can be cultivated through a variety of pedagogical approaches. Importantly, these approaches can be implemented via both digital and non-digital (i.e., unplugged) activities. For instance, non-digital activities could include stories and games while digital activities included programming which continues to be acknowledged as a crucial vehicle for developing and demonstrating CT skills. Drawing on the interconnected nature of cognitive, interpersonal, and interpersonal skills for deeper learning, K-12 teaching of CT also recognizes and promotes practices and mindsets, that must be cultivated, including tinkering, creating, collaborating, persevering, and debugging.

In recent years, with researchers designing learning experiences to teach AI/ML, it has become evident that students and teachers also need to understand the new data-driven paradigm of programming and computing. Shapiro, Fiebrink, and Norvig (2018) articulated these differences. They explained that the (traditional, pre-ML) computational model was driven by deterministic and logically verifiable algorithms that are central to the epistemology and practices of CS education. In contrast, ML models are characterized by an opaque composite of millions of parameters and an algorithm that is often not readable by humans. Where traditional software is built by human programmers who describe success as the accomplishment of a well-defined goal, "a typical ML system is built by describing the objective that the system is trying to maximize (what to achieve). The learning procedure then uses a dataset of examples to determine the model that achieves this maximization" (p. 27, Shapiro et al., 2018). Consequently, in addition to teaching students about how a computer executes a program and the pragmatic skills for writing and debugging programs that computers can execute, students now also need to build mental models about the statistical, data-driven, black-boxed model of computing. Tedre et al. (2021) called this *CT 2.0* and tabulated the distinctions between CT 1.0 and 2.0 as shown in Table 1.

**Table 1**: Paradigmatic Differences between CT 1.0 and CT 2.0 based on Tedre et al. (2021)

| CT 1.0 (Rule-Driven) | CT 2.0 (Data-Driven) |
| --- | --- |
| Formalize the problem | Describe the job and collect data from the intended context |
| Design an algorithmic solution | Filter and clean the data. Label the data |
| Implement a solution in a stepwise program | Train a model from the available data |
| Compile and execute the program | Evaluate and use the model |

CT 2.0's paradigm of programming necessitates building students' understanding of the shift from a top-down notion of the algorithm as a specific set of instructions for accurately reaching an end goal to a bottom-up, data-driven approach where patterns in a training dataset are discovered and decoded. In this latter approach, success is probabilistic, defined by an acceptable statistical confidence level. Results from such programs (or AI models, as they are commonly called) are not tested for accuracy through test cases or reaching a pre-defined outcome. Instead, they are evaluated or audited as acceptable or not based on running the model on a test dataset. Students thus must learn to appreciate the distinct nature of program evaluation and debugging in this new paradigm.

Additionally, current education efforts have been and must continue to embrace the increasing importance of data science in the age of "big data" (Lee, Wilkerson, & Lanouette, 2021), especially given that AI/ML is data-driven. This includes an increased understanding of data collection, curation, cleaning, analysis, and communication of ideas through visualizations and other means.

Each of these data science concepts has important situated and critical+ethical considerations, as discussed below.

While much of the paradigmatic change to CT 2.0 focuses on the data-centric nature of AI/ML, we cannot ignore that even machine learning applications use algorithms that guide their data selection. While these aspects tap into more difficult-to-understand mathematical concepts and practices, we should rise to the challenge of making them accessible to K-12 students, investigating what pedagogies make which ideas accessible at which grade band. As Broll and Grover (2023) demonstrated, learning can be designed to attend to through coding experiences in the NetsBlox block-based programming environment (Brady et al., 2022) that include both CT1.0 and CT2.0. For example, students explore data and its features through CoDAP, a dynamic data-exploration tool (cite), use the pre-ML paradigm to code a rule-based decision-tree classifier, and then work with parsons puzzles to code pieces of an ML training decision-tree model and run a pre-coded evaluation function to examine the accuracy of the model.

In the age of AI, we thus need a more expansive view of CT. Given that AI models are also computer programs—albeit ones following a different programming paradigm—CT+ must comprise both CT 1.0, CT 2.0, and an understanding of data and data practices. In such an expanded version, students learn about both paradigms—rule-driven and data-driven computational thinking—as part of their K-12 computing education. As part of a foundational understanding of computational problem-solving, students must develop an understanding of the pre-ML paradigm of automation and programming with CT 1.0. However, learning experiences involving AI/ML models and applications must also introduce students to CT 2.0. Furthermore, learning about these different programming paradigms also needs to include situated and critical dimensions, in addition to the cognitive, as Kafai & Proctor (2022) argued, leading us to call for CT+

## Situated

The call for CT+ also includes social-cultural dimensions of engaging in computational thinking and the creation of computational artifacts through emphasizing computational participation (Kafai and Burke, 2014) and computational action (Tissenbaum, Sheldon, & Abelson, 2019). These dimensions highlight the social side of computational thinking and afford learners opportunities to create and share their own computational artifacts as well as design apps and artifacts for social good. This shift from learning isolated computing concepts in a vacuum to designing personally relevant applications has in present days become a mainstay in most K-12 computing education initiatives (Kafai & Burke, 2014). This dimension further promotes learner agency and participation in computing for community and self as a means to develop a strong CS identity. Importantly, this situated dimension of CT+ acknowledges the socio-cultural contexts of learning—students' backgrounds and interests as well as the context of the classroom and the community where learning happens. This impacts the selection of relevant, real-world problems and contexts for AI/ML demonstrations, projects, and applications. Examples here include high school students designing dance (Payne et al., 2021), students examining social media data or engaging in cybersecurity activities (Broll & Grover, 2023).

The situated framing of CT also embraces an integration of CT into STEM and other school subjects. For instance, Weintrop and colleagues (2016) articulated a wider set of competencies to operationalize "CT for STEM" through a taxonomy that included data practices, modeling and simulation practices, computational problem-solving practices (including programming), and systems thinking practices. CT in its avatar as a generative problem-solving skill is heightened in contexts outside of CS; it is a critical cross-cutting 21st-century skill alongside creativity, collaboration, communication, and critical thinking, inspiring teachers of all subjects to incorporate CT in their teaching (Grover, 2021a). Importantly, data is a crucial linchpin in CT integration efforts, with CT being used to analyze or visualize data from a domain to answer questions or understand phenomena in the domain (Grover, 2021b).

Critical+Ethical+Justice-Centered

Furthermore, with CT+ we expand on the critical layer of the Kafai and Proctor (2022) framework to accentuate a focus on ethical and justice-centered considerations for computational designs to account for and address the enormous influence socio-political eco-systems and power structures wield on AI/ML models and developments writ large, which in turn impact students' lives and all of society. This includes attention to discriminatory data, biased AI designs and deployment practices, and application injustices leading to real-world patterns of inequity and discrimination. As students develop CT+ skills, they learn to be thoughtful about the provenance and fairness of the data they are training AI models on, what their criteria for success of the program are, who is impacted in various ways by the model, and how. CT+ learning designs need to incorporate culturally relevant pedagogies that also include a critical examination of societal disparities enabled by technology (Benjamin, 2019; Edouard, 2024).

Examples here include discussions on how sensor technologies that track body movements at home or in sports settings reveal issues of privacy and data control (Kumar & Worsley, 2023) or provide data scenarios in which students examine issues of algorithmic fairness (Salac & Ko, 2023). Furthermore, high school students could use algorithm auditing—a method of repeatedly checking the output of an algorithm to examine its function and possible bias—when they design and program machine learning applications (Morales-Navarro & Kafai, 2024). Other examples include high school teachers developing lessons for critical inquiry into the critical harm text-to-image generators create (Ali, Ravi, Moore, Abelson, & Breazeal, 2024) or engaging middle school students in identifying stakeholders' different perspectives in using AI systems (Dipaola, Payne, & Breazeal, 2022). While these cognitive, social, and critical+ethical+justice-centered dimensions could be developed and attended to individually, rich examples of CT+ pedagogy would attend to them in interconnected ways (e.g., White, 2024).

## DIRECTIONS FOR K-12 COMPUTING EDUCATION

The arrival of generative AI in the K-12 education landscape provides the impetus for the discussions educational researchers, teachers, administrators, parents, and students need to have on *who* should learn *what*, and *how*. A crucial recognition is that despite differences in the foci of topics and content taught, AI/ML is essentially a subdiscipline of CS, in that there is no AI without computers. We can build on lessons learned from introducing and integrating computing into the K-12 curriculum) and its associated challenges addressing equity and social justice (White, 2024) and leveraging Grover's (2024) guidance on teaching AI to K-12 students and teachers.

### Who

First, we need to focus on *who* should be involved, and this involves students, teachers, and other stakeholders. We need to focus on students, their backgrounds, interests, and social eco-systems to ensure identity development and empowerment as a key outcome. In particular, we need to incorporate equity-oriented, inclusive teaching strategies to reach learners from minoritized groups from the outset instead of adding them in as an afterthought. This is especially relevant to AI and machine learning as it has been shown to disproportionately adversely impact underserved populations and people of color (Benjamin, 2019). Furthermore, realizing a broad reach will require AI/ML to become part of teacher preparation and professional development. In the past decade, the field of K-12 CS education research has come to appreciate teachers' crucial role in the success of CSForAll efforts (DeLyser et al., 2018). Preparing teachers will be an equally crucial step. Here we must leverage learnings from research on teacher preparation for CS, for example, using unplugged activities as a starting point, connecting to concepts teachers are familiar with in their own subjects, building teachers' confidence, self-efficacy, and pedagogical content knowledge, co-designing curricular materials, creating teacher communities of practice, developing banks of adaptable resources, and attending to teachers' capacity for formative, classroom measurement of learning and feedback. Finally, we also need to educate various other stakeholders—such as parents and school and district administrators—on AI/ML so that they can help build supportive ecosystems in schools for K-12 education.

### What

A second focus should be on *what* should be learned. It is important that as we define the learning goals of CT+, we attend to all the dimensions of engagement and problem solving with computers and AI/ML. While we recognize that new concepts and practices need to be incorporated into CS frameworks and standards (CSK12, 2016), we also need to attend to aspects of CT that are relevant to AI/ML to a larger (such as understanding data and its features and pattern recognition) or lesser (such as algorithmic thinking) degree and developing robust mental models (or notional machines) of AI, address misperceptions and inaccurate stereotypes of the discipline to promote interest and better learning among ALL students. This will also require a comprehensive understanding of what students already know through their daily interactions with AI/ML applications and what appropriate learning progressions look like. Most importantly, we need to have an integrated focus on data, ethics, and algorithmic justice. Learners must understand that computational innovations are not value-neutral (Conrad 2022; Dipaola, Payne, & Breazeal, 2022). Many current efforts are overly focused on technical aspects with ethics only as an add-on (Sanusi, Oyelere, Vartiainen, Suhonen, & Tukiainen, 2023). We need to realize the social nature of data production (Lee, Wilkerson & Lanouette, 2021 ). With much of the attention given to machine learning applications in computational thinking 2.0, we should not forget that traditional or rule-based programming will not go away. While some programming practices will change as programmers can ask generative AI applications to create code, students will still need to be able to do the even more demanding task of checking whether the provided code completes the desired task.

### How

A third focus should be on *how* we should promote CT+. A key outcome of "Computer Science For All" research and development was the recognition that diverse approaches should be leveraged to achieve the initiative's cognitive and affective goals. Under this broad tent, many pedagogies flourished, focusing on introducing both teachers and learners to the new ideas of computational problem-solving in ways that were engaging, inclusive, and expansive. We must address the teaching of a technical discipline new to both teachers and students by leveraging this plurality of pedagogies including project-, inquiry-, and game-based learning, hands-on coding activities and explorations, and unplugged activities. Acknowledging the lingering disparities in computing and that broadening participation among minoritized communities has continued to be a challenge (White, 2024), we need to create inclusive learning environments that value all learners and motivate them to engage in AI as a subject in college or as a career (Tissenbaum, Weintrop, Holbert, & Clegg, 2021). It will be more crucial than ever to situate learning experiences in the context of the many, readily available, real-world contexts of relevance to students and their communities. Furthermore, we need to develop programming and modeling tools for students to engage with CT+. We have a 30-year history of successfully developing accessible tools to enable the development of CT 1.0. Tools like Scratch (Resnick et al., 2008) have made programming accessible to millions of young programmers around the globe. Low-floor, accessible block-based programming environments such as Snap! (Harvey & Monig, 2017) and NetsBlox (Brady et al., 2022) have already added new features that support data explorations and CT+. In addition, AI/ML sandboxes such as Teachable Machines (Carney et al., 2017) have made the creation of machine learning models accessible to young learners.. We need to develop more tools for K-12 learners that include features to work with data and AI models and help students build expansive CT competencies, both on the technical and ethical side.

## CONCLUSION

In this paper, we build on previous work defining (Grover & Pea, 2013) and framing (Kafai & Proctor, 2022) computational thinking for K-12 education. Defining and framing CT has been an ever-evolving journey as our understanding of learning and computing has grown along with the fast-paced growth of CS. The introduction of generative AI/ML necessitates an expansion of what we mean by CT. Hence our call for CT+. In line with previous framings, we call for a stronger emphasis on critical and ethical considerations in developing CT+ competencies. We also recommend several directions where efforts to promote CT+ should be focused, drawing on lessons learned from K-12 CS education during the last decades. Furthermore, we recommend that AI/ML topics and issues should not be limited to CS classes but follow efforts that have introduced CT in STEM topics and across the curriculum. While our focus was largely on computing & AI education, we acknowledge that teachers, administrators, parents, and other stakeholders need guidance on many other issues related to AI (Borasi et al., 2024). Realizing all goals and demands will require significant efforts and resources on the part of all stakeholders involved in the education ecosystem. CT+ could well aid in providing some of the answers in this monumental, multi-faceted, emergent enterprise.

11